\documentclass[aps,prb,twocolumn,showpacs]{revtex4}
\usepackage{graphicx}
\begin{document}
\draft

\title{A reassessment of the Burns temperature and its relationship
to the diffuse scattering, lattice dynamics, and thermal expansion
in the relaxor Pb(Mg$_{1/3}$Nb$_{2/3}$)O$_3$}

\author{P. M. Gehring,$^1$ H. Hiraka,$^{2,3}$ C. Stock,$^4$
S.-H. Lee,$^1$ W. Chen,$^5$ Z.-G. Ye,$^5$ S. B. Vakhrushev,$^6$ and
Z. Chowdhuri$^1$}
\address{$^1$NIST Center for Neutron Research, National Institute of Standards and Technology, Gaithersburg, Maryland 20899-6100, USA}
\address{$^2$Department of Physics, Brookhaven National Laboratory, Upton, New York 11973-5000, USA}
\address{$^3$Institute for Materials Research, Tohoku University, Sendai, 980-8577, Japan}
\address{$^4$ISIS Facility, Rutherford Appleton Laboratory, Chilton, Didcot, OX11 0QX, UK}
\address{$^5$Department of Chemistry, Simon Fraser University, Burnaby, BC, V5A 1S6, Canada}
\address{$^6$Ioffe Physical-Technical Institute, 26 Politekhnicheskaya, 194021 St. Petersburg, Russia}

\begin{abstract}
We have used neutron scattering techniques that probe time scales
from $10^{-12}$\,s to $10^{-9}$\,s to characterize the diffuse
scattering and low-energy lattice dynamics in single crystals of the
relaxor PbMg$_{1/3}$Nb$_{2/3}$O$_3$ from 10\,K to 900\,K. Our study
extends far below $T_c = 213$\,K, where long-range ferroelectric
correlations have been reported under field-cooled conditions, and
well above the nominal Burns temperature $T_d \approx 620$\,K, where
optical measurements suggest the development of short-range polar
correlations known as ``polar nanoregions" (PNR).  We observed two
distinct types of diffuse scattering. The first is weak, relatively
temperature independent, persists to at least 900\,K, and forms
bow-tie-shaped patterns in reciprocal space centered on $(h00)$
Bragg peaks.  We associate this primarily with chemical short-range
order.  The second is strong, temperature dependent, and forms
butterfly-shaped patterns centered on $(h00)$ Bragg peaks. This
diffuse scattering has been attributed to the PNR because it
responds to an electric field and vanishes near $T_d \approx 620$\,K
when measured with thermal neutrons. Surprisingly, it vanishes at
420\,K when measured with cold neutrons, which provide $\sim 4$
times superior energy resolution. That this onset temperature
depends so strongly on the instrumental energy resolution indicates
that the diffuse scattering has a quasielastic character and demands
a reassessment of the Burns temperature $T_d$.  Neutron
backscattering measurements made with 300 times better energy
resolution confirm the onset temperature of $420 \pm 20$\,K. The
energy width of the diffuse scattering is resolution limited,
indicating that the PNR are static on timescales of at least 2\,ns
between 420\,K and 10\,K. Transverse acoustic (TA) phonon lifetimes,
which are known to decrease dramatically for wave vectors $q \approx
0.2$\,~\AA$^{-1}$ and $T_c < T < T_d$, are temperature independent
up to 900\,K for $q$ close to the zone center.  This motivates a
physical picture in which sufficiently long-wavelength TA phonons
average over the PNR; only those TA phonons having wavelengths
comparable to the size of the PNR are affected. Finally, the PMN
lattice constant changes by less than 0.001\,\AA\ below 300\,K, but
expands rapidly at a rate of $2.5 \times 10^{-5}$\,1/K at high
temperature. These disparate regimes of low and high thermal
expansion bracket the revised value of $T_d$, which suggests the
anomalous thermal expansion results from the condensation of static
PNR.
\end{abstract}


\date{\today}
\maketitle

\section{Introduction}

The concept of nanometer-scale regions of polarization, randomly
embedded within a non-polar cubic matrix, has become central to
attempts to explain the remarkable physical properties of relaxors
such as PbMg$_{1/3}$Nb$_{2/3}$O$_3$ (PMN) and
PbZn$_{1/3}$Nb$_{2/3}$O$_3$
(PZN).~\cite{Ye_review,Park,Bokov_rev,Vug06:73}  The existence of
these so-called ``polar nanoregions'' (PNR) was first inferred from
the optic index of refraction studies of Burns and Dacol on PMN,
PZN, and other related systems,~\cite{Burns} and later confirmed
using many different experimental techniques including x-ray and
neutron diffraction,~\cite{Bonneau89:24,Mathan91:3,Zhao,Hirota06:75}
$^{207}$Pb NMR,~\cite{Blinc} and piezoresponse force
microscopy.~\cite{Shvartsman04:69}  Early small-angle x-ray
scattering and neutron pair distribution function (PDF) measurements
on PMN by Egami {\it et al}.\ cast doubt on the nano-domain model of
relaxors.~\cite{Egami}  However, the recent PDF analysis of Jeong
{\it et al}., which shows the formation of polar ionic shifts in PMN
below $\approx 650$\,K, and which occupy only one third of the total
sample volume at low temperatures, provides convincing support for
the existence of PNR.~\cite{Jeong05:94} Neutron inelastic scattering
data published by Naberezhnov {\it et al}.\ on PMN offered the first
dynamical evidence of PNR in the form of a prominent broadening of
the transverse acoustic (TA) mode that coincides with the onset of
strong diffuse scattering at 650\,K,~\cite{Naberezhnov,Koo} roughly
the same temperature ($\approx 620$\,K) as that reported by Burns
and Dacol in their optical study of PMN. This temperature, commonly
known as the Burns temperature, and denoted by $T_d$ in the original
paper, is widely viewed as that at which static PNR first appear.
Likely condensing from a soft TO mode, distinctive butterfly-shaped
and ellipsoidal diffuse scattering intensity contours centered on
$(h00)$ and $(hh0)$ Bragg peaks, respectively, are seen below $T_d$
in both PMN~\cite{Vakhrushev_JPCS,You,Xu_TOF} and PZN.~\cite{Xu06:5}
Similar diffuse scattering patterns were subsequently observed in
solid solutions of PMN and PZN with PbTiO$_3$ (PMN-$x$PT and
PZN-$x$PT); however these patterns appear only on the Ti-poor
(relaxor) side of the well-known morphotropic phase boundary
(MPB).~\cite{Mat06:74,Xu04:70} The polar nature of the strong
diffuse scattering, and thus its association with the formation of
PNR, was unambiguously established by several electric field studies
of PMN~\cite{Vakhrushev_efield,Stock:unpub} and
PZN-8\%PT,~\cite{Gehring_efield,Xu_memory,Xu_redistribution,Xu06:5}
all of which showed dramatic changes in the shape and intensity of
the diffuse scattering as a function of field strength and field
orientation. The Burns temperature $T_d$ thus represents what is
arguably the most important temperature scale in relaxors and is
several hundred degrees Kelvin higher than the critical temperature
$T_c$ (which for PMN is $\approx 210$\,K, but defined only in
non-zero electric field).

%
%
\begin{figure}
\includegraphics[width=3.0in]{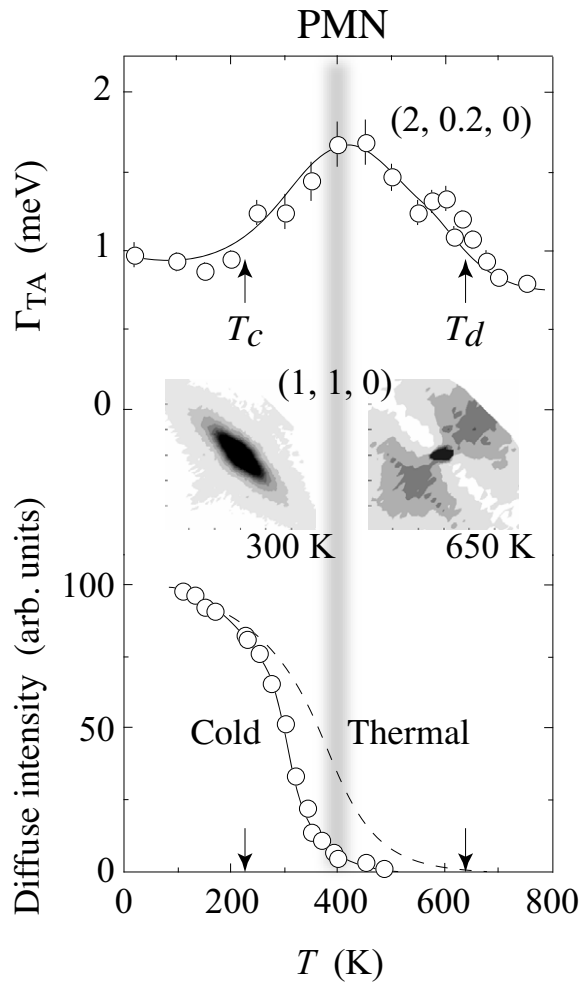}
\caption{\label{figure1} Top - temperature dependence of the TA
phonon energy width $\Gamma_{TA}$ measured with thermal neutrons by
Wakimoto {\it et al}.~\cite{Waki_sm} Middle - diffuse scattering
intensity contours measured at (110) below (300\,K) and just above
(650\,K) the Burns temperature $T_d$ using cold neutrons by Hiraka
{\it et al}.~\cite{Hiraka} Bottom - diffuse scattering temperature
dependence measured with cold neutrons (open circles) and thermal
neutrons (dashed line) by Hiraka {\it et al}.~\cite{Hiraka}}
\end{figure}
%
%

Recent studies of the neutron diffuse scattering in PMN by Stock
{\it et al}.,~\cite{Stock} and especially those by Hlinka {\it et
al}.~\cite{Hlinka_110} and Gvasaliya {\it et
al}.,~\cite{Gvasaliya,Gvasaliya2} have proven to be extremely
important because they demonstrated the utility of cold neutron
spectroscopy to the study of relaxors.  Generally speaking cold
neutrons are ill-suited to lattice dynamical studies because the
longer wavelengths necessarily limit their use to the study of
comparatively fewer (lower-$Q$) Brillouin zones.  On the other hand
cold neutron spectrometers provide significantly better energy
($\hbar \omega$) and wave vector ($q$) resolution than do their
thermal neutron counterparts. In addition, both the (110) and the
(100) Brillouin zones of PMN are accessible using cold neutrons with
wavelengths $\lambda \approx 4.26$\,\AA.  The combined cold and
thermal neutron study of PMN by Hiraka {\it et al}.\ exploited this
fact and uncovered several major new findings, two of which are
summarized in Fig.~\ref{figure1}.~\cite{Hiraka} The first finding is
that the temperature at which the strong diffuse scattering vanishes
depends on whether the measurement is made with cold or thermal
neutrons, i.\ e.\ the value of $T_d$ depends on the energy
resolution of the spectrometer. This fact, which is illustrated in
the bottom panel of Fig.~\ref{figure1}, indicates that the diffuse
scattering in PMN, and most likely other relaxors, contains a
substantial quasielastic component.  However, a consensus on this
issue is lacking; while the finding of Hiraka {\it et al}.\ is
consistent with the study of Gvasaliya {\it et
al}.,~\cite{Gvasaliya,Gvasaliya2} it contradicts that of Hlinka {\it
et al}.~\cite{Hlinka_110,eres} A second major finding is displayed
in the middle panel of Fig.~\ref{figure1} where intensity contours
of the diffuse scattering measured with cold neutrons around (110)
are shown to exhibit markedly different reciprocal space geometries
at 300\,K and 650\,K.  These data unambiguously demonstrate the
presence of two distinct types of diffuse scattering in PMN.  As the
PNR are absent at 650\,K $\ge T_d$, the physical origin of the
bow-tie-shaped diffuse scattering cross section is believed to come
primarily from the underlying chemical short-range order. We shall
refer to this as CSRO, although another commonly-used term for this
is compositionally/chemically ordered regions (COR).~\cite{Burton}

An intriguing perspective on the Burns temperature in PMN is
provided in the top panel of Fig.~\ref{figure1} where the TA phonon
energy width $\Gamma_{TA}$, which is inversely proportional to the
phonon lifetime, is plotted as a function of temperature for PMN at
the scattering vector $\vec{Q} = (2,0.2,0)$. These data were
measured with thermal neutrons by Wakimoto {\it et
al}.~\cite{Waki_sm} and are consistent with those of Naberezhnov
{\it et al}.\ in that the TA mode begins to broaden at $T_d \approx
620$\,K, the same temperature where the strong diffuse scattering
first appears.  These data also show that the TA broadening achieves
a maximum value (minimum lifetime) near 420\,K, which coincides with
the value of $T_d$ measured with cold neutrons. These data raise the
question of how to interpret the Burns temperature $T_d$ properly in
that they paint a picture, markedly different from that currently
held, in which the PNR are dynamic below $\sim 650$\,K and condense
into static entities only at a much lower temperature of 420\,K.

The goal of our study then is to determine the intrinsic value of
the Burns temperature $T_d$ and to clarify its relationship to the
diffuse scattering, lattice dynamics, and structure in PMN. To this
end we have carried out extensive measurements of the neutron
diffuse scattering cross section from 10\,K to 900\,K, far below
$T_c$ and well above the nominal value of $T_d$, that probe
timescales from $10^{-12}$\,s to $10^{-9}$\,s.   We find that a
300-fold improvement in energy resolution over that used by Hiraka
{\it et al}.,~\cite{Hiraka} obtained using neutron backscattering
techniques reproduces the same onset temperature for the diffuse
scattering; hence the intrinsic Burns temperature $T_d$ for PMN is
420\,K.  At the same time an enormous change in the thermal
expansion is observed near 420\,K, which is indistinguishable from
zero at low temperature. Given the revised value of $T_d$, this
result implies the existence of a direct influence of the PNR on the
intrinsic structural properties of PMN.

We also present new data on the effects of the PNR on the lattice
dynamics through measurements of the temperature and wave vector
dependence of the long-wavelength TA phonon energy width
$\Gamma_{TA}$ measured near (110) from 25\,K to 900\,K.  We find
that TA modes with reduced wave vectors $q \ll 0.2$\,\AA$^{-1}$
exhibit the same energy width at all temperature whereas those with
$q \approx 0.2$\,\AA$^{-1}$ exhibit a strongly temperature-dependent
broadening similar to that shown in the top panel of
Fig.\ref{figure1}.  This behavior contrasts with that observed in
thermal neutron studies of the TO mode, which exhibits a broadening
for all $q \le 0.2$\,\AA$^{-1}$. Previous neutron scattering work on
PMN-60\%PT by Stock {\it et al.},~\cite{Stock06:73} a material in
which there is no strong, temperature dependent diffuse scattering,
and thus no polar nanoregions, found no evidence of any TA phonon
broadening.  In this context, our data lend extremely strong support
to the PNR model: the lifetimes of TA modes with wavelengths
comparable in size to the PNR are strongly diminished by the PNR,
whereas long-wavelength (low $q$) TA phonons simply average over the
PNR and are unaffected.

\section{Experimental Details}

The neutron scattering data presented here were obtained using the
BT9 thermal neutron triple-axis spectrometer, the SPINS cold neutron
triple-axis spectrometer, and the cold neutron High Flux
Backscattering Spectrometer (HFBS), all of which are located at the
NIST Center for Neutron Research (NCNR).  On BT9, measurements of
the phonons and diffuse scattering were made at a fixed final
(thermal) neutron energy $E_f = $14.7\,meV ($\lambda = 2.36$\,\AA)
using the (002) Bragg reflection of highly-oriented pyrolytic
graphite (HOPG) crystals to monochromate and analyze the incident
and scattered neutron beams, respectively.  Horizontal beam
collimations were 40$'$-47$'$-S-40$'$-80$'$ (S = sample).  A
special, non-standard high $q$-resolution configuration was employed
to measure the thermal expansion in which the (004) Bragg reflection
from a perfect Ge crystal was used as analyzer and horizontal beam
collimations were tightened to 15$'$-47$'$-S-20$'$-40$'$.  The
choice of Ge was motivated by the close matching between the PMN
(022) (1.431\,\AA) and Ge (004) (1.414\,\AA) $d$-spacings, which
provides a significant improvement in the instrumental
$q$-resolution.~\cite{Xu_Acta}  On SPINS, which sits on the cold
neutron guide NG5, the phonon and diffuse scattering measurements
were made at a fixed final neutron energy $E_f=4.5$\,meV ($\lambda =
4.264$\,\AA) also using the (002) Bragg reflection of HOPG crystals
as monochromator and analyzer.  A liquid-nitrogen cooled Be filter
was located after the sample to remove higher order neutron
wavelengths from the scattered beam, and horizontal beam
collimations were set to guide-80$'$-S-80$'$-80$'$. The resultant
elastic ($\hbar \omega=0$) energy resolution for the SPINS
measurements was $\delta E = 0.12$\,meV half-width at half-maximum
(HWHM).

The High-Flux Backscattering Spectrometer was used to look for
dynamics that might be associated with the strong diffuse scattering
below $T_d$. This instrument uses a mechanically-driven Si(111)
monochromator to Doppler shift the energies of incident neutrons
over a narrow range centered about 2.08\,meV. Neutrons are
backscattered from the monochromator and proceed towards the sample
where they are scattered into a 12\,m$^2$ array of Si (111) crystals that
serve as analyzer. These neutrons are then backscattered a second
time by the analyzer, which selects the final neutron energy $E_f =
2.08$\,meV, into a series of detectors positioned about the sample.
The effective angular acceptance of each detector is $\approx
15^{\circ}$. The HFBS instrument is described in further detail
elsewhere.~\cite{Meyer03:74} The elastic energy resolution for the
HFBS measurements described here was $\delta E = 0.4$\,$\mu$eV
(HWHM).

Two high-quality single crystals of PMN, labeled PMN\,\#4 and
PMN\,\#5, were used in this study; both were grown using a
top-seeded solution growth technique.~\cite{Ye} The crystal growth
conditions were determined from the pseudo-binary phase diagram
established for PMN and PbO. The PMN\,\#4 and \#5 crystals weigh
2.7\,g (0.33\,cm$^3$) and 4.8\,g (0.59\,cm$^3$), respectively. At
300\,K the mosaic of each crystal measured at (220) is less
than 0.04$^{\circ}$ full-width at half-maximum (FWHM). Loss of PbO,
the formation of a pyrochlore phase, and the reduction of Nb$^{5+}$
are known to occur in PMN single crystals when subjected to high
temperatures under vacuum for extended periods of time.  This
process results in a dramatic blackening of the crystal, which is
normally of a transparent gold/amber color. While dielectric
measurements on such darkened crystals reportedly show little
difference from those on unheated samples,~\cite{Vakhrushev_private}
our measurements reveal a diminishment of the diffuse scattering
intensity after sustained and repeated heating. Therefore
experiments on the larger PMN crystal \#5 were limited to 600\,K or
less, while PMN crystal \#4 was used to obtain data above 600\,K.

Both samples were mounted with an [001] axis oriented vertically,
giving access to reflections of the form $(hk0)$.  For the high
temperature experiments, PMN crystal \#4 was wrapped in quartz wool,
mounted in a niobium holder secured by tungsten wire, and then
loaded into a water-cooled furnace capable of reaching temperatures
from 300\,K to 1800\,K. PMN crystal \#5 was mounted onto an aluminum
sample holder assembly placed inside an aluminum sample can, and
then loaded inside the vacuum space of a closed-cycle $^4$He
refrigerator that provides access to temperatures from 10\,K to
700\,K.  Each sample has a cubic lattice spacing of $a = 4.05$\,\AA\
at 300\,K; thus 1\,rlu (reciprocal lattice unit) equals $2\pi/a =
1.55$\,\AA$^{-1}$.

\section{Origins of the Diffuse Scattering: PNR versus CSRO}

%
%
\begin{figure}
\includegraphics[width=3.0in]{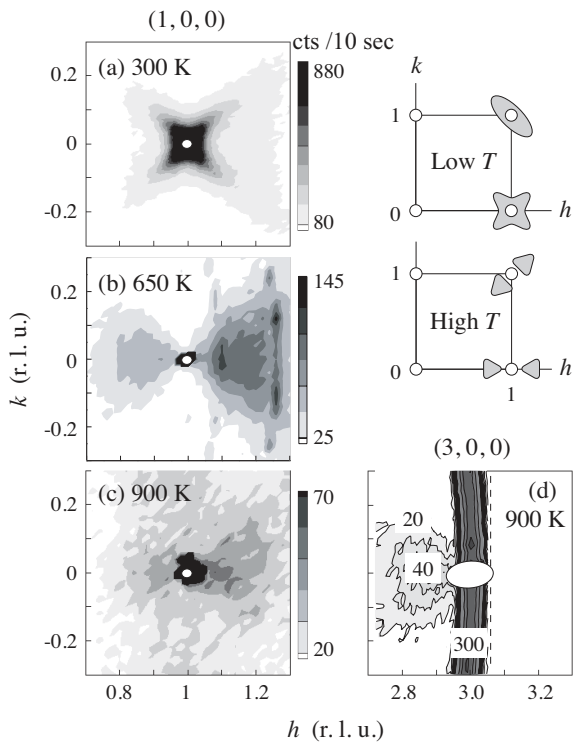}
\caption{\label{diffuse_contours} Diffuse scattering intensity
contours in PMN measured on SPINS near (100) are shown using a
linear gray scale at (a) 300\,K, (b) 650\,K, and (c) 900\,K. Diffuse
scattering intensity contours measured on BT9 at 900\,K near (300)
using thermal neutrons are shown in (d).  The data shown in panels
(a) and (b) are from Hiraka {\it et al}.~\cite{Hiraka} Schematic
diagrams of the diffuse scattering intensity contours below
(Low-$T$) and at $T_d$ (High-$T$) are shown in the upper right.}
\end{figure}
%
%

%
%
\begin{figure}[b]
\includegraphics[width=3.0in]{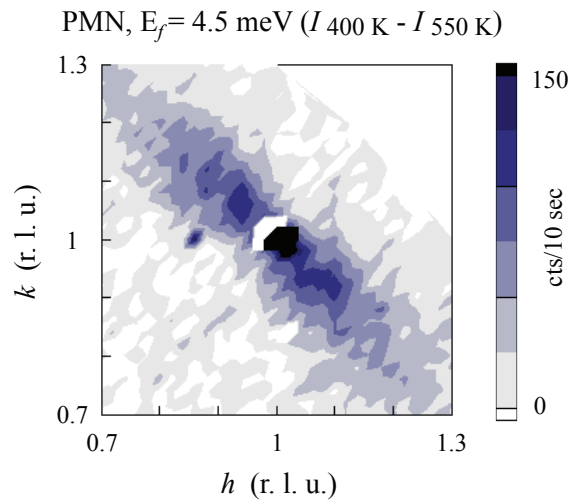}
\caption{\label{subtraction_figure} Contour plot of the difference
in diffuse scattering intensity near (110) measured at 550\,K and
400\,K. The result is the characteristic figure ``8'' pattern,
observed previously by Vakhrushev {\it et al}.~\cite{Vakhrushev}}
\end{figure}
%
%

The diffuse scattering in PMN crystal \#5 was studied by Hiraka {\it
et al}.~\cite{Hiraka} at and below the nominal Burns temperature
$T_d$ with good energy resolution ($\delta E \approx 0.12$\,meV
HWHM) using cold neutrons.  Data were measured in the (100) and
(110) Brillouin zones and are represented schematically in the upper
right portion of Fig.~\ref{diffuse_contours}.  The horizontal and
vertical axes in this figure correspond to the components of the
scattering vector $\vec{Q} = (hk0)$, which are measured in
reciprocal lattice units (rlu).  The ``Low-$T$'' ($T < T_d$) regime
is dominated by contributions from the PNR, where the diffuse
scattering intensity contours near (100) resemble a butterfly and
those near (110) resemble an ellipse for which the long axis is
oriented perpendicular to $\vec{Q}$.  This is shown in
Fig.~\ref{diffuse_contours}\,(a) where diffuse scattering data near
(100) at 300\,K reveal an intense butterfly-shaped
pattern.~\cite{Hiraka}  We note here that we see no evidence of any
``transverse component" to the diffuse scattering in any of our PMN
crystals like that reported previously by Vakhrushev {\it et
al}.~\cite{Sergey_NSE}  We assume that this component was due to the
crystal imperfections mentioned by these authors in their PMN
sample, which gave rise to a powder ring in their data at (100).

The same butterfly/ellipsoidal diffuse scattering geometry was shown
to persist in single crystals of PMN-$x$PT and PZN-$x$PT in studies
by, respectively, Matsuura {\it et al}.\ and Xu {\it et al}.\ for
compositions spanning the Ti-poor (relaxor) side of the morphotropic
phase boundary (MPB).~\cite{Mat06:74,Xu04:70} These results also
completely refute those of La-Orauttapong {\it et al}.\ who reported
that the orientation of the strong diffuse scattering varies with Ti
content in PZN-$x$PT and concluded that the PNR orientation changes
with doping.~\cite{Gop,comment}  For Ti-rich (tetragonal) PMN-$x$PT
compositions just beyond the MPB, Matsuura {\it et al}.\ found that
the strong, temperature dependent diffuse scattering vanishes and is
replaced by critical scattering.~\cite{Mat06:74} Matsuura {\it et
al}.\ also found that the $q$-integrated diffuse scattering
intensity increases with Ti content on the Ti-poor side of the MPB,
peaks near the MPB, then drops dramatically on crossing the MPB.
This finding is significant because it suggests that an intriguing
and direct correlation exists between the PNR and the piezoelectric
coefficient $d_{33}$, which exhibits the same dependence on Ti
content.~\cite{Park} A model based on pancake-shaped, ferroelectric
domains has been used successfully to fit the three-dimensional
diffuse scattering distributions measured in PZN-$x$PT with
high-energy x-rays.~\cite{Xu04:70,Welberry74:06,Welberry38:05} A
similar type of real-space structure has been proposed to explain
the diffuse scattering in the relaxor
KLi$_x$Ta$_{1-x}$O$_3$.~\cite{Waki74:06} On the other hand,
alternative models explaining the same diffuse scattering
distributions have also been proposed.~\cite{Pasciak}

In the ``High-$T$'' regime ($T \ge T_d$) there are no PNR, and the
associated butterfly-shaped diffuse scattering is no longer present.
The weak diffuse scattering that remains is thus argued to originate
primarily from the underlying chemical short range order (CSRO),
which reflects weak correlations between the Mg$^{2+}$ and Nb$^{5+}$
cations on the $B$-site of the perovskite $AB$O$_3$ structure. In
this regime the shapes of the diffuse scattering contours are
radically different, resembling a bow-tie in both $\vec{Q} = (h00)$
and $\vec{Q} = (hh0)$ Brillouin zones in which the diffuse
scattering extends mainly parallel to $\vec{Q}$. The only difference
between the contours near (100) and (110) appears to be in the
orientation of the triangular regions of diffuse scattering, which
point in towards (100), but away from (110).  Data taken near (100)
at 650\,K are displayed in
Fig.~\ref{diffuse_contours}\,(b).~\cite{Hiraka} At this temperature
the diffuse scattering intensities, shown using a linear gray scale,
are much weaker than those of the butterfly pattern at 300\,K.  We
further note that the intensities increase (become darker) from left
to right in panels (b) and (c), which corresponds to increasing $Q$.
Given the $(\vec{Q}\cdot\vec{u})^2$ dependence of the neutron
diffuse scattering cross section, this intensity signature implies
the presence of some short-range, correlated displacements $\vec{u}$
since otherwise, if $u = 0$, there would be no $Q$-dependence.  Thus
the weak diffuse scattering is not solely due to CSRO.

In Fig.~\ref{subtraction_figure} we plot the difference between the
diffuse scattering intensities measured near $\vec{Q} = (110)$ at
400\,K and 550\,K.  The resulting contours approximately reproduce
the ``figure 8'' pattern observed previously by Vakhrushev {\it et
al}.,~\cite{Vakhrushev}, and indicates that the ellipsoidal diffuse
scattering has a strong temperature dependence.  By contrast, the
bow-tie-shaped diffuse scattering is effectively subtracted out in
this analysis, which confirms that it has little to no temperature
dependence.  Thus the high-temperature diffuse scattering is not
associated with the Burns temperature $T_d$ or with the formation of
long-range polar correlations below $T_c$.  We note that such a
strictly incoherent treatment of the high-temperature (CSRO) and
low-temperature (PNR) components of the total diffuse scattering, as
described here, ignores the inevitable cross terms that must exist
between them.  On the other hand, if the high-temperature scattering
does arise primarily from CSRO, then it should be largely
independent of the ionic displacements that give rise to the
butterfly-shaped diffuse scattering (PNR) below $T_d$. The relative
weakness of the high-temperature diffuse scattering compared to that
at low temperatures also suggests such cross terms should be weak,
and this appears to be supported by the simple subtraction analysis
presented in Fig.~\ref{subtraction_figure} in that one effectively
recovers the ellipsoidal (not bow-tie) intensity contours.  For this
reason we believe it is a reasonable first approximation to treat
the two diffuse scattering components as being nearly independent.

%
%
\begin{figure}
\includegraphics[width=3.0in]{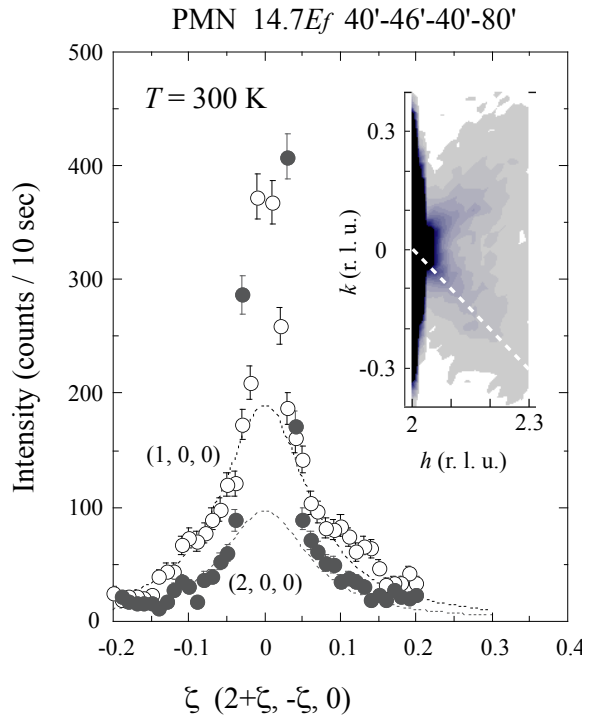}
\caption{\label{diffuse200} Diffuse scattering intensity measured on
BT9 at 300\,K as a function of reduced momentum transfer relative to
(100) (open circles) and (200) (solid circles).  The dotted lines
isolate the diffuse scattering component from the Bragg peak. The
inset shows the orientation of the scan relative to the familiar
butterfly pattern measured near (200) at 100\,K.}
\end{figure}
%
%

%
%
\begin{figure}
\includegraphics[width=3.0in]{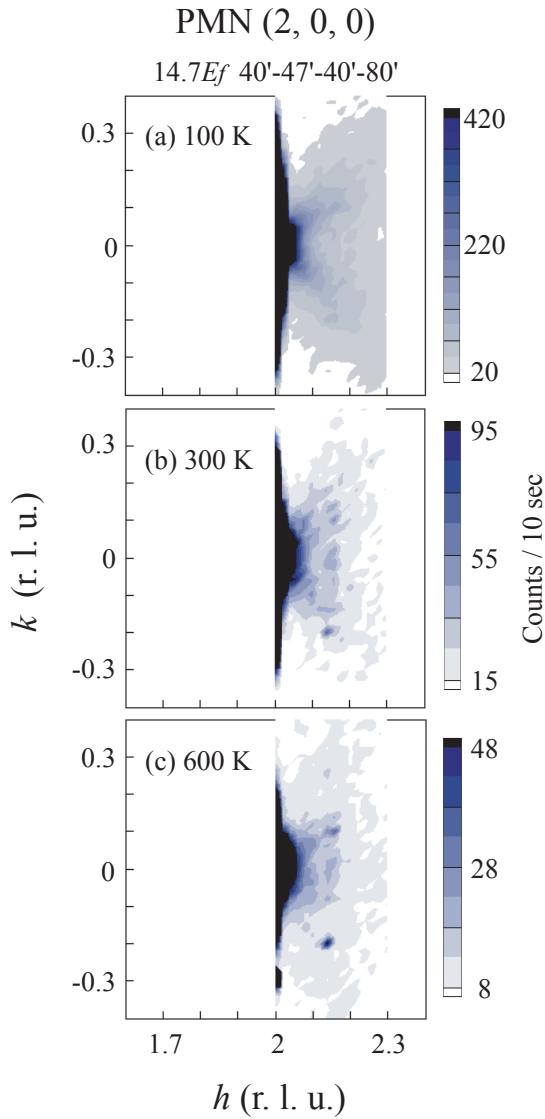}
\caption{\label{diffuse200_temp} Diffuse scattering contours
measured on BT9 near (200) are shown at (a) 100\,K, (b) 300\,K, and
(c) 600\,K.}
\end{figure}
%
%

We extended the diffuse scattering measurements to temperatures well
above $T_d$ using PMN crystal \#4, which we reserved for very high
temperature experiments.  Previous data taken on PMN and PZN
crystals heated to 1000\,K revealed significant evidence of sample
decomposition, so we limited our measurements to 900\,K.  Diffuse
scattering intensity contours at 900\,K for the (100) and the (300)
Brillouin zones are presented in Fig.~\ref{diffuse_contours}\,(c)
and \ref{diffuse_contours}\,(d), respectively.  Although the diffuse
scattering near (100) is quite weak, it is consistent with the
bow-tie geometry observed in Fig.~\ref{diffuse_contours}\,(b) at
650\,K.  We exploited the $Q^2$ dependence of the neutron diffuse
scattering cross section to obtain higher intensity by using thermal
neutrons to access (300). The (300) diffuse scattering intensity
contours are shown in Fig.~\ref{diffuse_contours}\,(d), where the
bow-tie pattern observed at 650\,K is still present at 900\,K. The
contours are truncated for $h > 3.02$\,rlu because of a mechanical
limit on the maximum scattering angle available on BT9, but the
triangular region on the low-$Q$ side of (300) is clearly evident.
That the bow-tie-shaped diffuse scattering persists to such high
temperature provides what is perhaps the most convincing evidence
that it arises mainly from CSRO.

During the course of our measurements we noticed that the diffuse
scattering in PMN crystal \#4 had diminished and was noticeably
weaker than that in PMN crystal \#5, which we had never exposed to
temperatures above 650\,K.  We also observed a broad ring of
scattering passing directly through $\vec{Q} = (300)$, which is
shown in Fig.~\ref{diffuse_contours}\,(d).  This feature was never
observed prior to heating this crystal to 900\,K.  Given the
length of time spent at high temperatures, this feature most likely
corresponds to a powder ring arising from partially decomposed
regions of PMN crystal \#4, which had turned entirely black after
exposure to high temperatures. These regions do not affect any other
data presented in this paper because the 900\,K measurements on PMN
crystal \#4 were the last ones performed on this sample. Therefore 
the powder ring appears only in Fig.~\ref{diffuse_contours}\,(d).

It is instructive to compare these results with those on PMN-60\%PT,
a composition that lies well beyond the morphotropic phase boundary
and undergoes a first-order ferroelectric transition from a cubic to
a tetragonal phase near 540\,K. An extensive study of this material
using neutron and high-energy x-ray scattering methods found no sign
of the strong, butterfly-shaped diffuse scattering at low
temperatures.~\cite{Stock06:73}  This result lends further support
to our association of the strong, temperature dependent diffuse
scattering with the PNR, which are absent in PMN-60\%PT.  Neutron
measurements on PMN-60\%PT do, however, reveal the presence of
bow-tie shaped diffuse scattering intensity contours at all
temperatures studied, which supports the identification of such
diffuse scattering with chemical short-range order between cations
on the $B$ site of the PMN perovskite $AB$O$_3$ structure.  This
picture is supported by theoretical work~\cite{Burton99:60} as well
as $^{93}$Nb NMR,~\cite{Laguta03:67} electron
microscopy,~\cite{Boul94:108} and polarized Raman
scattering~\cite{Svit03:68} measurements.  All of these studies
suggest that there is no temperature dependence to the bow-tie
shaped diffuse scattering below $\approx 1000$\,K, which is
consistent with our results on PMN over the extended temperature
range.

During our study of PMN we discovered that the diffuse scattering
near (200) is not as weak as previously
believed.~\cite{Vakhrushev,You,Hirota,Gop} To confirm this finding,
we made detailed measurements of the diffuse scattering intensity
near (200) at 300\,K and 100\,K along a trajectory in reciprocal
space that follows one wing of the butterfly intensity contour; this
is shown by the dashed line in the inset to Fig.~\ref{diffuse200}.
The results of the 300\,K scan are compared to an identical scan
measured in the (100) zone, both of which are shown in
Fig.~\ref{diffuse200}. These data demonstrate that the (100) diffuse
scattering cross section, represented by the dotted lines passing
through the open circles, is substantially larger than that at
(200), designated by the solid circles.  This result supports the
model of Hirota {\it et al}.\ in which the (unexpectedly) weak (200)
diffuse scattering cross section observed in PMN and other relaxors
can be explained by the presence of a uniform shift or displacement
of the PNR relative to the non-polar cubic matrix along the
direction of the local PNR polarization.~\cite{Hirota} Indirect
evidence for the existence of this shift has been obtained from
neutron scattering measurements of the anisotropic response of the
diffuse scattering in PZN-8\%PT to an electric field applied along
the [001] direction.~\cite{Gehring_efield}

Fig.~\ref{diffuse200_temp} shows diffuse scattering intensity
contours measured on BT9 at 100\,K, 300\,K, and 600\,K near(200);
these data illustrate that the (200) diffuse scattering intensity
follows the same temperature dependence as that measured in other
Brillouin zones, where the diffuse scattering is much stronger. As
the temperature is raised the diffuse scattering intensity decreases
in the same manner as that previously measured and observed in the
(100), (110), and (300) Brillouin zones.  This proves that the
diffuse scattering measured at (200) has the same origin as that in
other zones, i.\ e.\ that it is associated with the formation of
PNR.  At 600\,K in panel (c) one can already see the emergence of
the bow-tie-shaped diffuse scattering that is otherwise obscured by
the stronger PNR-related diffuse scattering at lower temperatures.
These data are important because they support the mode-coupling
analysis of Stock {\it et al}.~\cite{Stock}, which
assumes that the diffuse scattering in PMN in the (200) and
(220) Brillouin zones is much weaker than that in the (110) zone.
Thus we emphasize that while the neutron diffuse scattering
cross-section near (200) is not zero, it is small and consistent
with previous structure factor calculations.

\section{Diffuse Scattering Dynamics: A Reassessment of the Burns Temperature}

%
%
\begin{figure}
\includegraphics[width=3.75in]{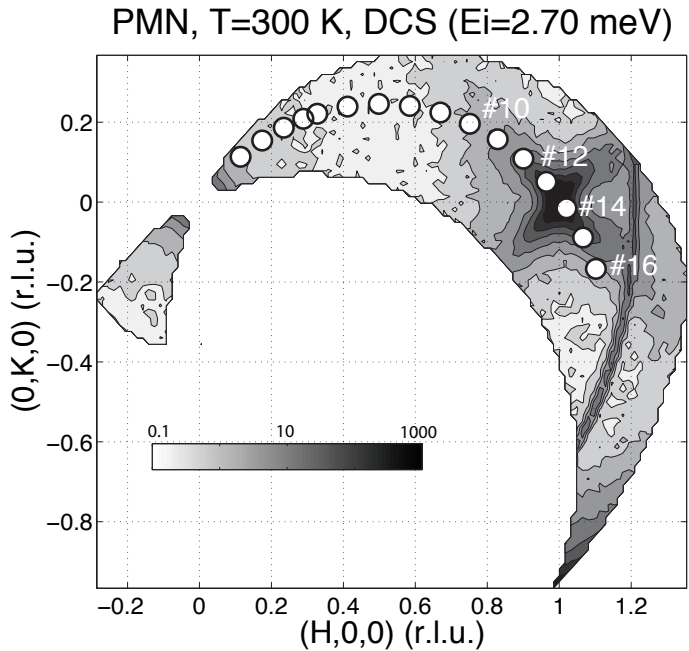}
\caption{\label{HFBS_det} Diffuse scattering intensity contours in
PMN measured on the NCNR Disk Chopper Spectrometer (DCS) near (100)
are plotted on a logarithmic gray scale at 300\,K.  Open circles
represent the reciprocal space locations of the detectors of the
High Flux Backscattering Spectrometer relative to the
butterfly-shaped intensity contours. Each open circle is a separate
detector, and the detector numbers are indicated. These DCS data are
taken from Xu {\it et al}.~\cite{Xu_TOF}}
\end{figure}
%
%

%
%
\begin{figure}
\includegraphics[width=3.75in]{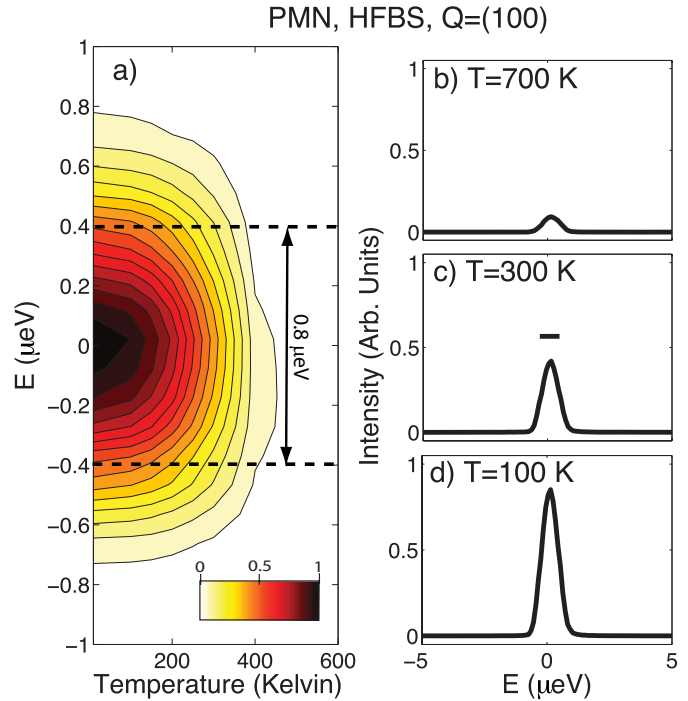}
\caption{\label{contour_figure}(Color online) (a) Contour plot of the
diffuse scattering intensity measured on the HFBS as a function of
energy transfer and temperature. Contours are shown on a linear
intensity scale; dashed lines indicate the full-width at
half-maximum (FWHM) of the peak linewidth at each temperature. Data
were summed over detectors 10-16 as illustrated in
Fig.~\ref{HFBS_det}. Panels (b), (c), and (d) show inelastic scans
at 700\,K, 300\,K, and 100\,K. The horizontal bar in panel (c)
represents the instrumental FWHM elastic energy resolution ($2\delta
E$).}
\end{figure}
%
%

The reciprocal space geometry of the strong diffuse scattering in
PMN was first characterized using x-ray diffraction and is
consistent with the neutron scattering data we have presented
here.~\cite{Vakhrushev_JPCS,You} The energy resolution provided by
x-ray diffraction ($\delta E \approx 1000$\,meV) is typically much
broader than that of thermal neutrons ($\delta E \approx 1$\,meV);
thus it was assumed that the strong diffuse x-ray scattering
originated from low-energy, soft, transverse optic (TO) phonons that
were captured by the large energy resolution.~\cite{You,Comes70:26}
However the cold neutron data of Hiraka {\it al}.\ provide a much
narrower elastic energy resolution of $\approx 0.12$\,meV HWHM and
show, unambiguously, that the diffuse scattering cross section
contains a component that is static on timescales of at least $\sim
6$\,ps below 420\,K as illustrated in Fig.~\ref{figure1}. This
result was subsequently confirmed on a separate PMN crystal by the
neutron study of Gvasaliya {\it et al}.\, which employed comparable
energy resolution.~\cite{Gvasaliya_JPhysC05}  Hence the observed
strong diffuse scattering cannot simply be the result of a soft,
low-lying TO phonon.  The TO mode must condense and/or broaden
sufficiently to produce the elastic diffuse scattering cross section
observed by Hiraka {\it et al}. Such a scenario is in fact suggested
by the corresponding thermal neutron data taken on BT9 using a
somewhat coarser energy resolution of $\approx 0.50$\,meV HWHM.  As
shown in Fig.~\ref{figure1}, an apparent elastic diffuse
scattering cross section is observed up to temperatures as high as
650\,K.

%
%
\begin{figure}
\includegraphics[width=3.25in]{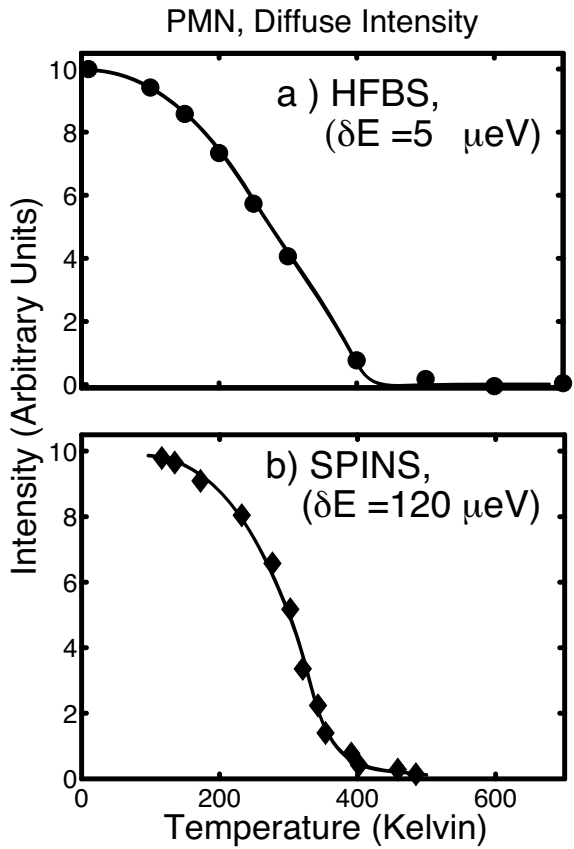}
\caption{\label{integrated} (a) Temperature dependence of the
diffuse scattering measured on the HFBS; data are integrated in
energy from $\pm$ 5 $\mu eV$ and in $Q$ over detectors 10-16 as
illustrated in Fig. \ref{HFBS_det}. (b) Temperature dependence of
the diffuse scattering intensity measured on SPINS at (1.05,0.95,0)
by Hiraka {\it et al}.~\cite{Hiraka} The SPINS energy resolution of
$\delta E$=120\,$\mu$eV HWHM provided the energy integration.}
\end{figure}
%
%

Motivated by these results, we looked for evidence of a dynamic
component of the strong diffuse scattering using the NCNR High-Flux
Backscattering Spectrometer (HFBS), which provides an elastic energy
resolution $\delta E = 0.4$\,$\mu$eV HWHM. We oriented PMN crystal
\#4 in the $(hk0)$ scattering plane, which is the same geometry used
for the triple-axis studies discussed in the previous sections.
Fig.~\ref{HFBS_det} displays the resulting locations of each of the
16 HFBS detectors relative to the butterfly-shaped diffuse
scattering pattern at (100) measured previously using the NCNR Disk
Chopper Spectrometer (DCS).~\cite{Xu_TOF} From this figure it
can be seen that detectors 10 through 16 sample different parts of
the wings of the strong diffuse scattering.  In particular,
detectors 13 and 14 lie close to the (100) Bragg peak. Because the
instrumental $Q$-resolution of the HFBS is relatively poor, we
checked for the presence of Bragg contamination in the integration
analysis by removing contributions from detectors 13 and 14.  This
did not change any of the results. Moreover the PMN study by
Wakimoto {\it et al}.\ showed the (202) Bragg peak intensity changes
by less than 10\%\ between 50\,K and 300\,K.~\cite{Waki_coupling}
Therefore, in the following analysis we have integrated the
intensity from detectors 10 through 16.

The energy dependence of the diffuse scattering as a function of
temperature is illustrated in Fig.~\ref{contour_figure}.  Panel (a)
shows a color contour plot of the peak linewidth in energy as a
function of temperature after subtraction of a high temperature
background. This background was taken to be the average of the
intensity measured at 600\,K and 700\,K because no strong diffuse
scattering is observed with either cold neutron triple-axis or
neutron backscattering methods at these temperatures.  The distance
between the dashed lines represents the measured energy width
(full-width at half-maximum) as a function of temperature.  Typical
energy scans are displayed in panels (b), (c), and (d) at 700\,K,
300\,K, and 100\,K, respectively.  We see that the peak centered at
$\hbar \omega = 0$ is equal to the instrumental energy resolution
$\delta E = 0.4$\,$\mu$eV HWHM at all temperatures.  Based on this
analysis, we conclude that the strong diffuse scattering is elastic
on timescales of at least $\tau \approx \hbar/\delta E \sim 2$\,ns.
Our results are thus consistent with the neutron-spin echo study on
PMN of Vakhrushev {\it et al}.~\cite{Sergey_NSE}

Fig.~\ref{integrated}\,(a) shows the temperature dependence of the
diffuse scattering intensity integrated over $\pm 5$\,$\mu$eV. These
data are compared to those measured on PMN crystal \#5 using the
cold neutron spectrometer SPINS, which are plotted in panel (b).
That the temperature dependences from the backscattering and SPINS
measurements agree within error in spite of a 300-fold
(120\,$\mu$eV/0.4\,$\mu$eV) difference in energy resolution proves
that static, short-range polar order first appears at much lower
temperatures than has been understood from previous data measured
with much coarser energy resolution.  These data thus demand a
revised value for the Burns temperature: $T_d = 420 \pm 20$\,K. We
mention, however, that if the diffuse scattering energy width obeys
an Arrhenius or some power law form, which has been suggested for
spin glasses in Ref.~\onlinecite{Som82:25}, then a more detailed
analysis based on data taken closer to the onset of the diffuse
scattering at 420\,K will be required to confirm (or reject) the
existence of such alternative dynamic contributions to the diffuse
scattering energy width.

To gain a better understanding of the apparent quasielastic nature
of the diffuse scattering, we examined the temperature dependence of
the low-energy transverse acoustic (TA) modes in greater detail.  In
particular we focused on measurements of the temperature dependent
broadening of the transverse acoustic phonon over a range of reduced
wave vectors $q$ that approach the zone center ($q=0$). These data
are discussed in the following section.

\section{TA and TO Phonons: Effects of the PNR}

\begin{figure}
\includegraphics[width=3.25in]{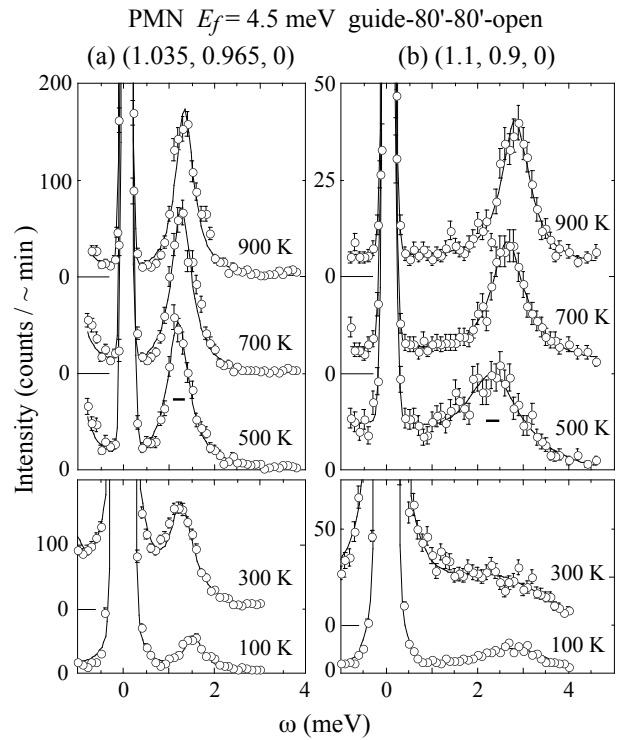}
\caption{\label{TA_SPINS} TA phonon lineshapes in PMN measured on
SPINS with cold neutrons at $\vec{Q} = (1.035,0.965,0)$ and
(1.1,0.9,0) from 100\,K to 900\,K. }
\end{figure}

An extensive series of inelastic measurements were made on both PMN
crystals \#4 and \#5 in the (110) Brillouin zone using the SPINS
spectrometer in an effort to map out the temperature dependence of
the TA mode for reduced wave vectors $q \ll q_{wf}$ and $q \approx
q_{wf}$, where $q_{wf} = 0.14$\,rlu $\approx 0.2$\,\AA$^{-1}$ is the
wave vector associated with the so-called ``waterfall'' effect, and
below which the long-wavelength, soft TO modes in PMN, PZN, and
PZN-8\%PT are observed to broaden markedly at temperatures below
$T_d$.~\cite{Gehring_pzn8pt,Gehring_Aspen,Waki_sm,Gehring_sm,Gehring_pzn}
This zone was chosen because the TA phonon dynamical structure
factor for (110) is much larger than that for (100).  The TA phonon
energy lineshapes were studied at two very different values of the
reduced wave vector $q$, measured relative to the (110) zone center,
for temperatures between 100\,K and 900\,K. These data are presented
in Fig.~\ref{TA_SPINS}. The data shown on the left-hand side of this
figure correspond to $\vec{Q} = (1.035, 0.965, 0)$ or $q = \sqrt{2}
\cdot 0.35$\,rlu = 0.05\,rlu = 0.077\,\AA$^{-1}$, which is less than
half of $q_{wf}$.  These data show that the TA lineshape at this
small wave vector remains sharp and well-defined at all temperatures
and has an intrinsic energy width that is larger than the
instrumental resolution (shown by the small, solid horizontal bar at
500\,K). By contrast, the data on the right-hand side of
Fig.\ref{TA_SPINS} correspond to $\vec{Q} = (1.1, 0.9, 0)$ or $q =
\sqrt{2} \cdot 0.10$\,rlu = 0.14\,rlu = 0.22\,\AA$^{-1}$, which is
nearly equal to $q_{wf}$. In this case it is quite evident that the
TA lineshape broadens dramatically, becoming increasingly
ill-defined below $T_d$, especially at 300\,K, but then sharpens at
lower temperature, e.\ g.\ at 100\,K. This behavior is the same as
that observed for the soft, zone-center TO mode in PMN by Wakimoto
{\it et al}.\ measured at (200).~\cite{Waki_sm}

%
%
%
%

\begin{figure}
\includegraphics[width=3.25in]{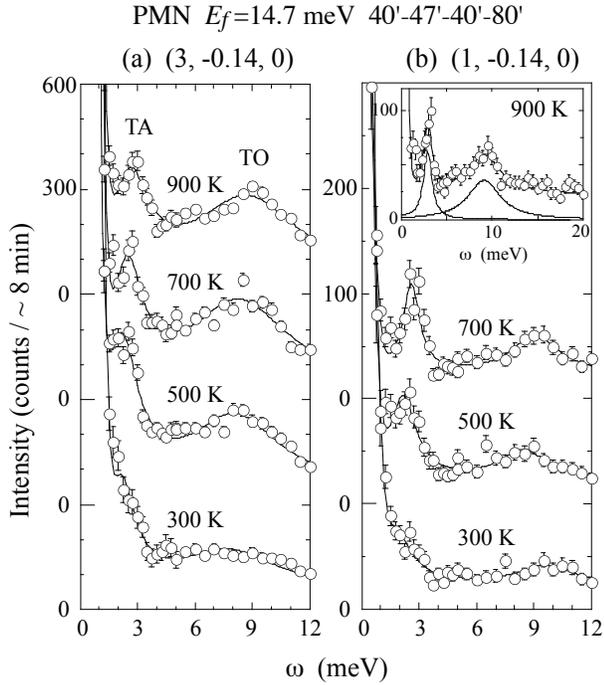}
\caption{\label{figure_100_300} TA and TO phonon lineshapes in PMN
measured on BT9 with thermal neutrons near the waterfall wave vector
at $\vec{Q} = (3,- 0.14,0)$ and (1,-0.14,0) from 300\,K to 900\,K. }
\end{figure}

In addition to these cold neutron measurements of the TA mode using
SPINS, other measurements were made using thermal neutrons on BT9
to characterize the TO mode lineshapes for reduced wave vectors $q
\approx q_{wf}$ from 300\,K to 900\,K.  In these experiments data
were taken using a water-cooled furnace with PMN crystal \#4 in both
the (300) and (100) Brillouin zones for $q = q_{wf} = 0.14$\,rlu and
are shown in Fig.\ref{figure_100_300}.  As can be seen on the
left-hand side of this figure, both the TA and TO modes are
well-defined at 900\,K.  The scattering intensity below 1-2\,meV
increases sharply because of the comparatively coarser energy
resolution intrinsic to BT9 ($\approx 0.5$\,meV HWHM), which uses
thermal neutrons, compared to that on SPINS.  Even so, the TA mode
is easily seen at both 900\,K and 700\,K. However at 500\,K the TA
mode is less well-defined. This occurs in part because the mode has
broadened, but also because the low-energy scattering has increased,
which results from the onset of the strong diffuse scattering due to
the PNR.  At 300\,K the TA mode is nearly indistinguishable from the
sloping background, and the TO mode has become significantly damped.
One also sees a gradual softening of the TO mode from 900\,K down to
300\,K that is of order 1-2\,meV.  The same data were taken in the
(100) zone, which appear in the right-hand portion of
Fig.~\ref{figure_100_300}. Essentially the same trends are observed
in this zone, with the TA mode again becoming almost
indistinguishable from background at 300\,K.  One difference is that
the TA mode is better separated at all temperatures from the steep
increase in scattering seen at low energies. These data are
consistent with the conjecture of Stock {\it et al}.~\cite{Stock}
that the low-energy TA phonon is coupled to the diffuse scattering
centered around the elastic position. The inset on the top panel on
the right-hand side displays the full TA-TO phonon spectrum at
900\,K out to 20\,meV.

%
%
%
%
%
%
\begin{figure}
\includegraphics[width=3.0in]{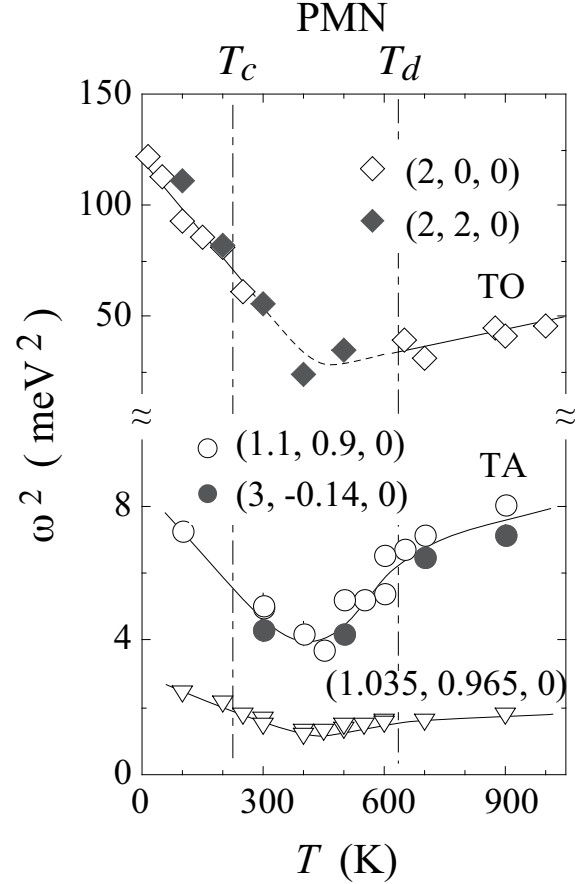}
\caption{\label{energy_squared} The square of the phonon energy is
plotted versus temperature for the zone-center TO modes measured at
(200) (open diamonds) and (220) (solid diamonds), taken from
Wakimoto {\it et al}.\ and Stock {\it et al}.~\cite{Waki_sm,Stock},
as well as for three TA modes measured at (1.035,0.965,0) (open
triangles), (1.1,0.9,0) (open circles), and (3,-0.14,0) (solid
circles).  Linear behavior is consistent with that expected for a
conventional ferroelectric soft mode.}
\end{figure}
%
%

The square of the energies of both the TA and TO modes presented in
the two previous figures are plotted versus temperature in
Fig.~\ref{energy_squared}. This is done to draw attention to the
similarity between the low-energy lattice dynamics of PMN and the
behavior expected for a conventional ferroelectric, for which the
soft mode frequency squared $\omega^2$ varies linearly with
$(T-T_c)$. Data for the zone-center soft mode measured at (200) and
(220) have been taken from Wakimoto {\it et al}.\ and Stock {\it et
al}.\ and added to the top of Fig.~\ref{energy_squared} for ease of
comparison.~\cite{Stock,Waki_sm} The data taken by Wakimoto {\it et
al}.\ are based on energy scans measured at constant-$Q$ at the zone
center, whereas Stock {\it et al}.\ determined the zone-center TO
energies by extrapolating from data obtained at non-zero $q$.  The
extrapolation technique permits phonon energies to be extracted in
the temperature range where the zone-center TO mode is heavily
damped. These values for the zone-center TO phonon energy have been
confirmed by infrared spectroscopy.~\cite{Bovtun04:298} What this
figure immediately reveals, then, is a surprising ferroelectric
character of the zone center TO mode above $T_d$ and below $T_c$,
and a corresponding change in the behavior of the TA mode energy
that is bracketed by the same two temperature scales, which are
denoted by the two vertical dash-dotted lines.  We note that no such
change in the acoustic phonons was observed in PMN-60\%PT where PNR
and the associated strong diffuse scattering are absent; therefore
the softening of the TA mode is directly related to the development
of PNR.~\cite{Stock06:73}  A further interesting feature is the
minimum that is present in all of these data near 420\,K, which is
the same temperature at which the strong diffuse scattering first
appears when measured with high energy resolution, namely the
revised value of the Burns temperature. Therefore, the onset of the
diffuse scattering is directly associated with the softening of the
TO mode; this is yet further evidence that it is associated with the
formation of static, short-range polar correlations.

It is very important to note that the square of the TA phonon energy
measured at $\vec{Q}$=(1.1,0.9,0), which corresponds to $q \sim
0.14$\,rlu, shows a much more pronounced minimum at 420\,K than does
that measured at $\vec{Q}$=(1.035,0.965,0), which corresponds to $q
\sim 0.05$\,rlu. This shows that long-wavelength TA phonons exhibit
a much weaker response to the formation of static short-range, polar
correlations. This can be understood in terms of a simple physical
picture in which those phonons with wavelengths comparable to the
size of the PNR are strongly affected (damped) by the PNR whereas
longer wavelength phonons simply average over the PNR and are thus
not affected by the presence of static, short-range polar
correlations. This idea was previously proposed to explain the
anomalous damping of the TO mode for wave vectors $q \le q_{wf}$
near the zone center.~\cite{Gehring_pzn8pt} However, no strong
diffuse scattering is seen in PMN-60\%PT and thus no PNR are
present, even though the anomalous TO broadening is still observed;
hence this TO broadening, which gives rise to the waterfall effect,
cannot be associated with the presence of static, short-range polar
correlations. On the other hand, the idea that the acoustic phonons
are affected by PNR is confirmed by the absence of any such acoustic
phonon broadening in PMN-60\%PT. Thus PNR have a significant effect
on the low-energy acoustic phonons over a limited range of reduced
wave vectors that may be related to the size of the PNR. In light of
the diffuse and inelastic scattering data that have been analyzed so
far, we now turn to the detailed measurement of the thermal
expansion in PMN.

\section{Thermal Expansion:  Invar-like Effect Below $T_d$}

\begin{figure}
\includegraphics[width=3.0in]{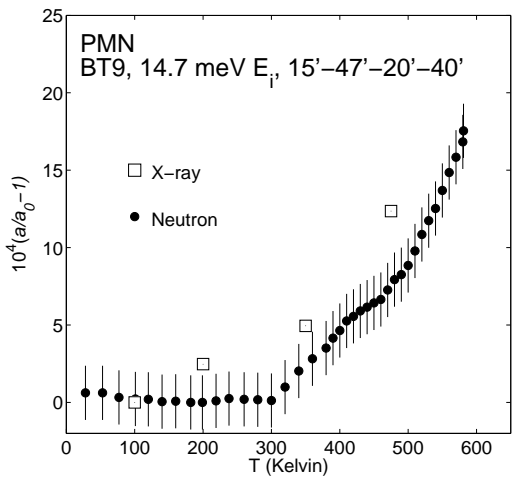}
\caption{\label{thermal_expansion} Lattice strain of single crystal
PMN derived from the (220) Bragg reflection measured from 30\,K to
580\,K on BT9. Data were obtained on heating using a perfect Ge(004)
analyzer, an incident neutron energy $E_i = 14.7$\,meV, and
horizontal beam collimations of 15$'$-47$'$-S-20$'$-40$'$ to obtain
extremely high $q$-resolution. X-ray data from Dkhil {\it et al}.\
are indicated by the open squares for comparison.~\cite{Dkhil}.}
\end{figure}

A variety of experimental techniques have been used to measure the thermal expansion in both single crystal and powder samples of PMN, which includes x-ray diffraction, laser dilatometry, and neutron diffraction.~\cite{Shebanov,Arndt,Dkhil} Here we present extremely high $q$-resolution neutron measurements ($\delta q \approx 0.0005$\,rlu = 0.0008\,\AA$^{-1}$ HWHM) of the cubic unit cell lattice parameter $a$ of PMN crystal \#5 using the special triple-axis configuration described in the experimental section.
This configuration, which employs a perfect single crystal of Ge as analyzer, also provides an exceptional energy resolution $\delta E \approx 5 - 10$\,$\mu$eV HWHM.  The resulting data are plotted in Fig.~\ref{thermal_expansion} in terms of the strain $a/a_0 - 1$, where $a_0$ is the lattice parameter at 200\,K. Measurements were made on heating from 30\,K to 580\,K using a closed-cycle $^4$He refrigerator after having first cooled to 30\,K. X-ray data measured by Dkhil {\it et al}.~\cite{Dkhil} on a single crystal of PMN, shown as open squares, are provided for comparison.

%
%
%
%

Several features are interesting to note. First, at low temperature the system exhibits an invar-like effect in which the cubic lattice parameter changes by less than 0.001\,\AA; indeed, the data below 320\,K are consistent with a thermal expansion of zero.  At higher temperatures, however, the average thermal expansion is $2.5 \times 10^{-5}$\,1/K. That these disparate regions of null and high rates of thermal expansion bracket the revised value for the Burns temperature suggests that a direct connection exists between the onset of static, short-range polar correlations and the structural properties of PMN.  This behavior seems to be consistent with that of PZN, which also exhibits an increase in the thermal expansion at temperatures above that where the diffuse scattering first appears.~\cite{Agrawal87:34}

There is ample evidence of similar behavior reported by other groups in samples of PMN and PMN-$x$PT. A low-temperature invar-like effect was observed in single crystal PMN-10\%PT, where a transition to a high rate of thermal expansion was found at 400\,K; the thermal expansion for this sample at high temperature is $1 \times 10^{-5}$\,1/K, which is very close to that measured here.~\cite{Gehring_pmn10pt} The x-ray study of ceramic samples of PMN by King {\it et al}.\ also show a transition between low and high rates of thermal expansion, but larger values for both.~\cite{King} X-ray and neutron work conducted by Bonneau {\it et al}.\ on PMN powders yielded onset temperatures and values for the thermal expansion consistent with our single crystal measurements.~\cite{Bonneau91:91} Finally, an invar effect was also observed in the laser dilatometry study by Arndt and Schmidt on a ceramic sample of PMN, which covered a range from 300\,K to 800\,K.~\cite{Arndt}

Even though there is a general trend towards a larger thermal expansion for temperatures above that where the diffuse scattering
is onset, there is some sample dependence and some differences between powders and single crystals.  As noted by Ye {\it et
al}.,~\cite{Ye10pt} powder measurements of PMN yield a different slope for the thermal expansion measurements than do those of Dkhil {\it et al}.~\cite{Dkhil}  Also, studies using neutron strain scanning techniques found different thermal expansion coefficients as a function of depth in single crystal samples.~\cite{Conlon04:70,Xu05:74}  Therefore, part of the discrepancy observed between different samples may be associated with surface effects.  In this regard, we also note the presence of a change in the slope of the strain curve between 400\,K and 500\,K; since this change is not observed in other PMN studies, we believe this feature to be sample-dependent and thus extrinsic.

The phase-shifted model of polar nanoregions proposed by Hirota {\it et al}.\ provides a plausible starting point from which to understand the anomalous invar-like behavior in PMN below the Burns temperature $T_d$. This model is based on the observation that the Pb, Mg, Nb, and O ionic displacements obtained from the diffuse scattering measurements of Vakhrushev {\it et al}.~\cite{Vakhrushev} on PMN can be decomposed into a center-of-mass conserving component, consistent with the condensation of a soft, transverse optic mode, and a uniform scalar component, corresponding to an acoustic phase shift. In so doing Hirota {\it et al}.\ were able to reconcile the discrepancies between the structure factors of the diffuse scattering and the soft TO modes and, in particular, to explain the weakness of the diffuse scattering intensity observed in the vicinity of the (200) Bragg peak.~\cite{Hirota} The idea that the PNR are uniformly displaced with respect to the underlying cubic lattice in a direction parallel to the PNR polarization has already been used by Gehring {\it et al}.\ to explain the persistence of the strong diffuse scattering in a single crystal of PZN-8\%PT against large electric fields applied along [001].~\cite{Gehring_efield} In that study it was shown that the diffuse scattering near (003), which measures the projections of the ionic displacements along [001], decreases under an external [001] field as expected.  By contrast, the diffuse scattering near (300), which measures the projections of the ionic displacements along [100], i.\ e.\ perpendicular to the field direction, remains unaffected even for field strengths up to 10\,kV/cm. Thus a uniform polar state is not achieved. This surprising behavior can be understood if one assumes that the electric field applied along [001] removes or reduces the PNR shifts along [001] while preserving those along [100]. The diffuse scattering should then decrease anisotropically with field, as is observed.  In the context of the invar-like behavior of PMN below $T_d$, we speculate that such uniform shifts of the PNR could effectively stabilize the lattice against subsequent thermal expansion at lower temperature. Such a scenario is consistent with the fact that both PMN-10\%PT and PMN-20\%PT retain cubic structures down to low temperature when examined with high $q$-resolution neutron diffraction methods.~\cite{Gehring_pmn10pt,Xu68:03}

\section{Discussion and Conclusions}

We have presented a comprehensive neutron study of the diffuse scattering in the relaxor PMN.  We have greatly extended the timescales of previous neutron measurements by taking data on three different spectrometers, BT9, SPINS, and the HFBS, which provide elastic energy resolutions of 500\,$\mu$eV, 120\,$\mu$eV, and 0.4\,$\mu$eV HWHM, respectively.  While the backscattering data represent a 300-fold improvement in energy resolution over those obtained by Hiraka {\it et al}.~\cite{Hiraka} on SPINS, both yield an onset temperature of $420 \pm 20$\,K for the diffuse scattering. This indicates that the PNR in PMN are static on timescales of at least 2\,ns below 420\,K, but are apparently dynamic at higher temperatures. Thus the true Burns temperature in PMN, which was originally interpreted by Burns and Dacol as the condensation temperature of static regions of local, randomly-oriented, nanometer-scale polarization,~\cite{Burns} is $420 \pm 20$\,K, not 620\,K.  Independent evidence of the existence of purely dynamic, short-range, polar correlations has been reported in PMN-55\%PT, a composition with no strong diffuse scattering and thus no static PNR, by Ko {\it et al}.\ who observed phenomena that are typically related to the relaxation of dynamic PNR.~\cite{Ko}  These include significant Brillouin quasielastic scattering, a softening of the longitudinal acoustic phonon mode, and a deviation of the dielectric constant from Curie-Weiss behavior over an 80\,K temperature interval above $T_c$.

Previous measurements of the temperature dependence of the neutron diffuse scattering have been extended in this study to 900\,K, well above $T_d$. In so doing we have unambiguously established the existence of two types of diffuse scattering based on the observation of two markedly different temperature dependencies, one of which vanishes at $T_d$ and one of which does not. We associate the strong, temperature-dependent, diffuse scattering with the formation of static, short-range, polar correlations (PNR) because of its well documented response to an external electric field, and because it first appears at the same temperature at which the soft (polar) TO mode reaches a minimum in energy.  We associate the weak, temperature-independent, diffuse scattering, which shows no obvious change across either $T_c$ or $T_d$, with chemical short-range order because it persists to extremely high temperature.  We further confirm the observations of Hiraka {\it et al}.,~\cite{Hiraka} who first characterized the distinctly different reciprocal space geometries of both types of diffuse scattering, and we show that the bow-tie shape of the weak diffuse scattering persists up to 900\,K $\gg T_d$.

Key effects of the strong, temperature-dependent, diffuse scattering, and thus of the PNR, on the low-energy lattice dynamics of PMN are also highlighted in this study. The neutron inelastic measurements on PMN-60\%PT by Stock {\it et al}.\ prove conclusively that PNR cannot be the origin of the anomalous broadening of long-wavelength TO modes observed in PMN, PZN, and other compounds, also known as the waterfall effect, because PMN-60\%PT exhibits the same effect but no strong diffuse scattering (no PNR).~\cite{Stock06:73}  By contrast, many studies have shown that PNR do broaden the TA modes in PMN ~\cite{Naberezhnov,Koo,Stock} and in PZN-$x$PT,~\cite{Gop} but that such effects are absent in compositions with no PNR such as PMN-60\%PT.~\cite{Stock06:73} Our cold neutron data show that these effects are $q$-dependent. Whereas long-wavelength TA modes with reduced wave vectors $q \ll 0.2$\,\AA$^{-1}$ remain well-defined and exhibit a nearly constant energy width (lifetime) from 100\,K to 900\,K, shorter wavelength TA modes with reduced wave vectors $q \approx 0.2$\,\AA$^{-1}$ broaden substantially, with the maximum broadening occurring at $T_d = 420$\,K. This result motivates a very simple physical picture in which only those acoustic phonons having wavelengths comparable to the size of the PNR are significantly scattered by the PNR; acoustic modes with wavelengths much larger than the PNR are largely unaffected because they simply average over the PNR.  Models describing this effect have been discussed elsewhere.~\cite{Stock} In particular, very recent work by Xu {\it et al}.\ has revealed the presence of a phase instability in PZN-4.5\%PT that is directly induced by such a PNR-TA phonon interaction. This is shown to produce a pronounced softening and broadening of TA modes in those zones where the diffuse scattering is strong, and provides a natural explanation of the enormous piezoelectric coupling in relaxor materials.~\cite{Xu_Nature}

In addition, we have performed neutron measurements of the thermal expansion with extremely high $q$- and $\hbar \omega$-resolution over a broad temperature range extending well below $T_c$ and far above $T_d \sim 420$\,K. In agreement with many other studies, our single crystal samples of PMN exhibit little or no thermal expansion below $T_d$, behavior that is reminiscent of the invar effect, but an unusually large thermal expansion coefficient of $2.5 \times 10^{-5}$\,1/K above $T_d$, where the strong diffuse scattering is absent. The crossover between null and large coefficients of thermal expansion coincides closely with $T_d$, which suggests that the appearance of static PNR strongly affects the thermal expansion in PMN and thus provides a structural signature of the formation of short-range, polar correlations in zero field. The model of uniformly displaced, or phase-shifted, PNR proposed by Hirota {\it et al}., which successfully predicts the anisotropic response of the strong diffuse scattering to an electric field, offers a simplistic, yet plausible, framework in which to understand this anomalous behavior.

Finally, it is satisfying to note that the revised value of $T_d = 420 \pm 20$\,K is consistent with the dielectric susceptibility data of Viehland {\it et al}., from which a Curie-Weiss temperature $\Theta = 398$\,K was derived.~\cite{Viehland} Such good agreement between $T_d$ and $\Theta$ solidifies our identification of the strong diffuse scattering with the condensation of the soft TO mode, which reaches a minimum frequency at $T_d = 420$\,K.  However this begs the question of how one should interpret the original Burns temperature of $\sim 620$\,K.  At present there are two broadly divergent opinions on this issue, one of which considers $\sim 620$\,K to be a meaningful temperature scale in PMN, and one of which does not.  As it turns out, this debate is closely tied to another on how many temperature scales are needed to describe the physics of relaxors.  We offer no final resolution to this discussion.  Instead, we close our paper with a brief summary of the primary studies supporting these contrasting points of view, which is by no means comprehensive, so that the readers may draw their own conclusions.

A number of experimental studies report evidence of either structural or dynamic changes in PMN in the temperature range 600\,K -- 650\,K, starting with the optical index of refraction measurements of Burns and Dacol.~\cite{Burns}  Siny and Smirnova were the first to observe strong, first-order Raman scattering in PMN at high temperatures, which, being forbidden in centrosymmetric crystals, implied the presence of some type of lattice distortion.~\cite{Siny}  In 2002 Vakhrushev and Okuneva calculated the probability density function for the Pb ion in PMN within the framework of the spherical layer model using x-ray powder diffraction data.~\cite{Vakhrushev_2002} It was shown that this probability density evolves from a single Gaussian function centered on the perovskite $A$-site to a double Gaussian form between 635\,K and 573\,K, and that the positions of maximum density for the lead ion follow a power law $(T_d - T)^{0.31}$ with $T_d = 635$\,K.  This picture was developed further by Prosandeev {\it et al}.\ within a model that ascribed the changes in the lead probability density function and subtle variations in the thermal expansion near 620\,K to a crossover from soft-mode to order-disorder dynamics.~\cite{Prosandeev} In 2004 Egami {\it et al}.\ first proposed that at very high temperatures the vibrations of the oxygen octahedra are sufficiently faster than those of the Pb ions that, on average, a high-symmetry, local environment is obtained, whereas at temperatures below $\sim 680$\,K the Pb and O ions become dynamically correlated.~\cite{Egami_2004}  This led to the subsequent reinterpretation of the Burns temperature ($\sim 600$\,K) as being the point below which dynamic PNR first form, which was based on the dynamic PDF measurements of Dmowski {\it et al}.\ obtained with an elastic energy resolution of 4\,meV at six temperatures (680\,K, 590\,K, 450\,K, 300\,K, 230\,K, and 35\,K).~\cite{Dmowski}  Interestingly, after integrating their data from -5\,meV to +5\,meV, Dmowski {\it et al}.\ saw evidence of a third temperature scale in PMN of order 300\,K, and thus distinct from both $T_c \sim 210$\,K and $T_d \sim 600$\,K, which they associated with the formation of static PNR.  Similar ideas have very recently been put forth by Dkhil {\it et al}.\, based on extremely interesting acoustic emission and thermal expansion measurements, and also in a general summary written by Toulouse.~\cite{Dkhil_3T,Toulouse}

A different approach was taken by Stock {\it et al}.\, who proposed a unified description of the lead-based relaxors on the basis of striking similarities between PMN and the Zn-analogue PZN, both of which display strong, temperature-dependent, diffuse scattering; identical soft mode dynamics; yet no long-range, structural distortion at low temperature in zero field.~\cite{Stock_PZN,Xu_PZN}  Arguing by analogy with magnetic systems, Stock {\it et al}.\ considered a three-dimensional Heisenberg model with cubic anisotropy in the presence of random fields in which the Heisenberg spin corresponds to the local ferroelectric polarization, the cubic anisotropy represents the preferential orientation of the polarization along the <111> axes, and the isotropic random magnetic field corresponds to the randomly-oriented local electric fields that originate from the varying charge of the $B$-site cation.  Following a suggestion by Aharony, Stock {\it et al}.\ considered the case where the Heisenberg term dominates over the random field term, and the cubic anisotropy term is the weakest of the three.  In this picture there would be just two distinct, static, temperature scales.  For $T > T_d$, the cubic anisotropy is irrelevant and therefore the system should behave like a Heisenberg system in a random field. In this case the excitation spectrum is characterized by Goldstone modes and therefore no long-range order is expected in the presence of random fields.~\cite{Aharony} Instead the system forms polar nanoregions in a paraelectric background. The second temperature scale $T_c$ appears at low temperatures where the cubic anisotropy term becomes important, and in this limit the system should resemble an Ising system in the presence of a random field.  This model thus explains the local ferroelectric distortion characterized by the recovery of the soft-optic mode and, although a 3D Ising system in equilibrium should display long-range order in the presence of small random fields, as is observed in magnetic systems, nonequilibrium effects with long time scales become dominant.  The presence of such nonequilibrium effects may explain the lack of long-range ferroelectric order in PMN and PZN as well as the history dependence of physical properties such as the linear birefringence.  The phase-shifted nature of the polar nanoregions may also create another energy barrier against the ordered phase at $T_c$.  Stock {\it et al}.\ are also able to explain the physics of compounds beyond the MPB, such as PMN-60\%PT, within this model, for which only one temperature scale ($T_c$) exists. As the Ti concentration Ti increases, the relaxor phase diagram crosses over to a ferroelectric regime, and this can be understood as an increase in the strength of the cubic anisotropy term that is simultaneously accompanied by a reduction of the random fields as must occur in the limit of pure PbTiO$_3$.~\cite{Stock06:73} It should be mentioned that other studies have invoked random fields to explain the static and dynamic properties of relaxors.~\cite{Westphal,Pirc,Fisch}

Independent of the validity of either of the two pictures summarized above, the seminal finding of our study of remains that the strong diffuse scattering in PMN first appears at a temperature that depends sensitively on the energy resolution of the measurement.  This fact inevitably raises interesting questions about the significance of the previous value of the Burns temperature ($\sim 620$\,K).  If the strong diffuse scattering in PMN results from the soft TO mode, then other techniques based on x-ray diffraction, thermal neutron scattering, or neutron PDF, which provide comparatively much coarser energy resolution, will be unable to discriminate between low-energy, dynamic, short-range polar correlations and truly static ($\hbar \omega=0$) PNR because any low-energy, polar correlations will be integrated into the elastic channel by the broad energy resolution.  Thus as the TO mode softens, the associated low-energy, polar, spectral weight will fall within the energy resolution at temperatures above $T_d = 420$\,K, and the net result will be an artificially higher value of $T_d$ that increases with the size of the energy resolution.  This effect should be especially pronounced for phonon modes that are broad in energy (i.\ e.\ that have short lifetimes), and this is clearly the case for the soft TO mode in PMN, which exhibits a nearly overdamped lineshape and a minimum frequency precisely at $T_d = 420$\,K.  While it is true that the structure factor of the strong diffuse scattering is inconsistent with that of the soft TO mode, the phase-shifted model of Hirota {\it et al}.\ provides a way to reconcile this discrepancy.  In this respect, we simply suggest that the previous value of $T_d \sim 620$\,K might not represent a physically meaningful temperature scale in PMN.  Future studies examining the values of $T_d$ in other relaxor compounds with improved energy resolution are thus of interest.

Our reassessment of the Burns temperature $T_d$ immediately clarifies an intimate relationship between the formation of static, short-range polar correlations and the consequent effects on both the low-energy lattice dynamics and structure of PMN.  Cold neutron triple-axis and backscattering spectroscopic methods conclusively show the existence of static, short-range polar correlations, only below $T_d = 420 \pm 20$,K on timescales of at least 2\,ns.  Thermal neutron measurements of the lattice dynamics reflect the presence of these static PNR through the presence of a distinct minimum in both the soft TO and TA mode energies, both of which occur at 420\,K.  The effect of PNR on the lattice dynamics is evident only for TA modes having wave vectors of order 0.2\,\AA$^{-1}$, a fact that could be exploited to determine the size of the PNR.  At the same time an enormous change in the coefficient of thermal expansion is seen near $T_d$, below which the crystal lattice exhibits invar-like behavior.

\section{Acknowledgments}

We would like to thank A.\ Bokov, Y.\ Fujii, K.\ Hirota, D.\ Phelan, S.\ Shapiro, S.\ Wakimoto, and G.\ Xu for stimulating discussions. This study was supported in part by the U.\ S.\ - Japan Cooperative Neutron-Scattering Program, the Natural Sciences and Engineering Research Council (NSERC) of Canada, the National Research Council (NRC) of Canada, the Japanese Ministry of Monbu-Kagaku-shou, RFBR grants 08-02-00908 and 06-02-90088NSF, the U.\ S.\ Dept.\ of Energy under contract No.\ DE-AC02-98CH10886, the U.\ S.\ Office of Naval Research under grant No.\ N00014-06-1-0166, and by the NSF under grant No.\ DMR-9986442.  We also acknowledge the U.\ S.\ Dept.\ of Commerce, NIST Center for Neutron Research, for providing the neutron scattering facilities used in this study, some of which are supported in part by the National Science Foundation under Agreement No. DMR-0454672.


\begin{references}
\bibitem{comment} These authors show schematic diffuse scattering intensity contours in Fig.~4 of their paper (no data are shown) for PZN-4.5\%PT and PZN-9\%PT that are wrong because they do not possess the required crystal lattice symmetry.
\end{references}
\end{document}